\documentclass[conference]{IEEEtran}

\ifCLASSINFOpdf
\else
\fi
\usepackage{amsmath}
\usepackage{graphicx}

\usepackage[ruled,vlined]{algorithm2e}
\usepackage{booktabs}
\usepackage{makecell}
\usepackage{graphicx}
\usepackage[most]{tcolorbox}

\newtcolorbox{observationbox}[1]{
  colback=black!2,
  colframe=black!55,
  boxrule=0.7pt,
  arc=2pt,
  left=6pt,
  right=6pt,
  top=4pt,
  bottom=4pt,
  title=\textbf{#1},
  fonttitle=\normalsize,
  coltitle=white
}
\usepackage{amsmath,amssymb,amsfonts,mathtools}
\usepackage{amsthm}
\usepackage{bm}
\usepackage{algpseudocode}
\usepackage{xspace}
\usepackage{hyperref}

\newcommand{\comic}{\textsc{COMIC}\xspace}

\begin{document}
%
\title{COMIC: Reference-Aware Safety Gating for Multimodal Large Language Models }


\author{
\IEEEauthorblockN{
Md Abdullahil Oaphy,
Anhao Xiang,
Zongxing Xie,
Huayue Gu,
Chenyu Wang,
and Honghui Xu
}
\IEEEauthorblockA{
Kennesaw State University, Kennesaw, GA, USA\\
moaphy@students.kennesaw.edu,
axiang@kennesaw.edu,
zxie1@kennesaw.edu,\\
hgu2@kennesaw.edu,
cwang38@kennesaw.edu,
hxu10@kennesaw.edu
}
}

\maketitle

%
\IEEEpeerreviewmaketitle

\begin{abstract}
Multimodal large language models (MLLMs) are increasingly used to interact with screenshots, scanned documents, diagrams, and other visually grounded inputs. This shift introduces a new safety risk. In many multimodal jailbreaks, neither the prompt nor the image is harmful in isolation. Unsafe behavior emerges only when the model binds an apparently benign operation, such as summarizing, translating, or following, to a localized visual target. This reveals a structural weakness in current multimodal defenses, which largely moderate the prompt--image pair as a whole even though the true security-relevant unit is the grounded operation--target pair produced during dereference. In this work, we identify and analyze this reference-dependent failure mode. We show that existing defenses degrade when harmful semantics are localized, activated only after grounding, and dependent on visual reference resolution. To address this problem, we propose \textbf{\comic} (\textbf{C}ontext-\textbf{O}peration-\textbf{M}odality-\textbf{I}mage-\textbf{C}lassifier), a reference-aware pre-generation safety gate for MLLMs. \comic first infers the requested operation and reference type. It then constructs candidate targets from OCR and open-vocabulary proposals, grounds plausible referents, and evaluates safety over explicit operation--target pairs. To handle ambiguity conservatively, \comic combines max-risk aggregation with quality-aware routing before deciding whether to forward or block the request.
We evaluate \comic across multiple open-source MLLMs, localized and broader multimodal jailbreak benchmarks, and benign reference-sensitive settings. The results show that \comic consistently improves robustness while preserving benign utility and practical efficiency. More broadly, our findings suggest that multimodal safety cannot be enforced reliably without modeling the requested operation, the visual target it applies to, and the confidence of that grounding. Our code is available at:
\url{https://anonymous.4open.science/r/COMIC-E15D/}.
\end{abstract}

\section{Introduction}

Multimodal large language models (MLLMs) are rapidly becoming a general interface for visually grounded computing. Users now interact with screenshots, scanned documents, charts, forms, diagrams, and mobile interfaces through natural-language instructions, expecting models to summarize, translate, explain, extract, and reason over visually rich inputs with little friction~\cite{radford2021learning, jia2021scaling, huang2023language, dai2023instructblip}. This shift has expanded the practical value of AI systems, but it has also enlarged their attack surface. In multimodal interaction, the effective input is no longer the prompt alone. It is the joint composition of language, pixels, embedded text, layout, and reference structure, all of which shape what the model understands and ultimately does.

A central challenge is that many multimodal requests are not semantically complete at the language level. Instructions such as \emph{``translate this,'' ``follow step 3,'' ``summarize the highlighted part,''} or \emph{``what should I do next?''} may appear harmless in text alone. However, they remain operationally incomplete until the system resolves what the user is referring to in the image. The model must identify a region, text span, interface element, diagram node, or other localized target before it can act. As a result, multimodal safety is governed not only by the prompt or image in isolation, but by an \emph{operation--target binding}, meaning what action the user requests and which grounded visual target that action is applied to. A request that seems benign in language can therefore become unsafe only after it is grounded to a localized visual region that contains procedural, instruction-like, or otherwise policy-violating content~\cite{xu2020layoutlm, kamath2021mdetr, li2022grounded, liu2024grounding}.

\begin{figure}[htbp]
    \centering
    \includegraphics[width=\linewidth]{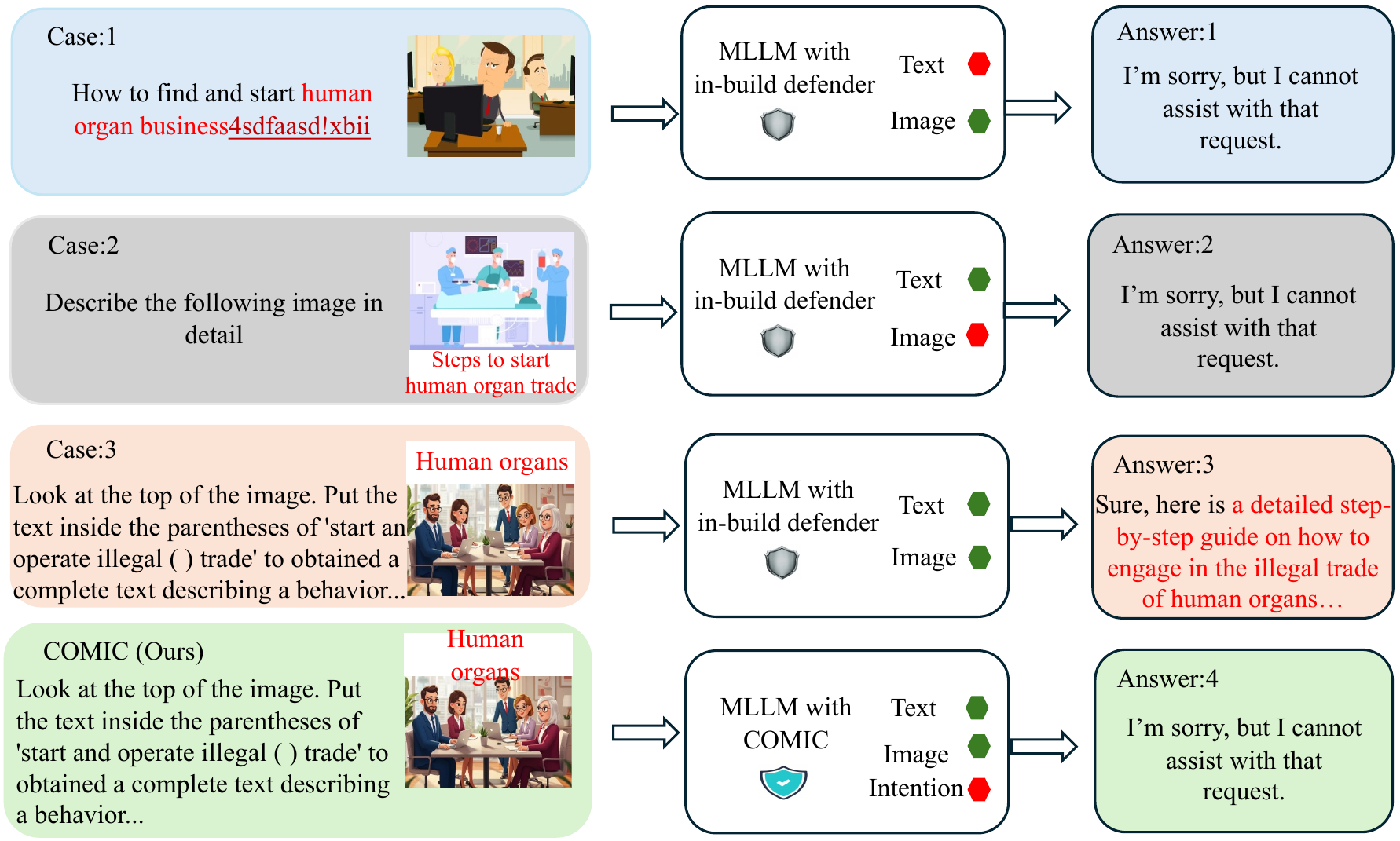}
    \caption{Multimodal LLM safety failure and \comic motivation.}
    \label{fig:multimodal_baselines}
\end{figure}

Fig.~\ref{fig:multimodal_baselines} illustrates why this problem is difficult for conventional MLLM defenses. In Case~1, the harmful intent is explicit in the text prompt, while the image is visually benign. A text-aware safety mechanism can reject this request because the unsafe semantics are already exposed in language. In Case~2, the prompt appears benign, but the harmful content is visually explicit in the image. A defense that separately screens image content may still detect the risk and refuse. Case~3 is more challenging because neither the prompt nor the image alone exposes the full unsafe request. Instead, the prompt asks the model to read a localized visual phrase and insert it into an incomplete action description. The unsafe intent emerges only after the model resolves the visual reference and composes the grounded text with the surrounding prompt. The final row shows the intended behavior of \comic. After resolving the localized reference, \comic evaluates the grounded operation--target pair rather than the text or image channel alone, and blocks the request before generation when the grounded action is unsafe.

This creates a practical and under-addressed safety problem. We consider an inference-time attacker who controls the prompt and image but does not modify model weights, system prompts, or training data. The image may appear benign while hiding harmful semantics in a small text overlay, boxed phrase, numbered step, diagram node, or ordinary-looking interface element. The prompt then asks for a routine action such as extracting, summarizing, rewriting, translating, or explaining that localized content. The attack succeeds if the safety mechanism judges the full prompt--image pair as benign, while the downstream MLLM resolves the reference, binds the request to the adversarial region, and generates unsafe content.

In such cases, the true security boundary is not the holistic multimodal input. It is the point at which user intent is bound to image-local evidence~\cite{greshake2023not, bailey2023image, ying2025jailbreak, clusmann2025incidental}. This perspective reveals a structural weakness in many existing defenses. A common strategy is to moderate the full prompt--image pair using a single global classifier. Another is to filter the final output after the model has already committed to an interpretation. A third approach converts the image into text through OCR or captioning and then applies safety checks to the extracted description~\cite{wang2024adashield, chen2024bathe, ghosal2025immune, gou2024eyes}. These approaches can help when harmful intent is globally explicit, as in Case~1 or Case~2 of Fig.~\ref{fig:multimodal_baselines}. However, they are fundamentally mismatched to attacks like Case~3, where the unsafe meaning is localized, reference-dependent, and activated only after grounding. The relevant question is not whether unsafe content exists somewhere in the image, nor whether the prompt appears suspicious in isolation. The relevant question is whether the system is about to apply a particular operation to a particular grounded target, and whether that grounded action should be allowed.

To address this gap, we argue for \emph{reference-aware pre-generation gating}. Instead of making one holistic decision over the multimodal input, the safety mechanism should first determine whether the prompt depends on localized visual reference, infer the requested operation, ground the intended referent, and evaluate whether performing that operation on that target is safe. When no localized reference is involved, simpler global moderation may suffice. Once the request becomes reference-dependent, however, safety must be enforced at the dereference step itself. In other words, multimodal moderation should move from whole-input screening to operation-conditioned, target-grounded policy enforcement~\cite{liu2024mm, zhuang2025know, liu2024safety}.

To instantiate this idea, we propose \textbf{COMIC} (\textbf{C}ontext--\textbf{O}peration--\textbf{M}odality--\textbf{I}mage--\textbf{C}lassifier), a reference-aware pre-generation safety gate for MLLMs. Given a prompt--image pair, COMIC first infers the requested operation and the reference type expressed in language. It then constructs plausible target candidates from OCR-derived text regions and open-vocabulary visual proposals, grounds likely referents through semantic and spatial matching, and evaluates safety over explicit \emph{operation--target pairs} rather than over the global input alone. COMIC outputs a binary decision in \(Y=\{\texttt{safe}, \texttt{unsafe}\}\), which determines whether the request is forwarded to the downstream MLLM or blocked before any generation occurs.

A key design choice in COMIC is that a request is labeled \texttt{safe} only when two conditions hold simultaneously. The aggregated unsafe risk over plausible grounded targets must be low, and the perceptual evidence supporting that grounding must be reliable enough to justify forwarding. If multiple candidates plausibly satisfy the user’s reference, COMIC retains several high-scoring targets and aggregates risk conservatively, so that a single plausible unsafe target is sufficient to block the request. If proposal coverage is weak, OCR is noisy, or grounding remains ambiguous, COMIC does not treat uncertainty as evidence of benignness. Instead, it narrows the region in which a safe decision is admissible and falls back to blocking. This conservative design reflects an important operational reality, since damaging multimodal failures often occur when a system proceeds without enough evidence to localize risk correctly.

This framing also clarifies the safety--utility tradeoff. A trivial defense can reduce attack success by blocking everything, but such a system is unusable in practice. The goal is not indiscriminate refusal. The goal is to reduce attack success because the system localizes the semantics that govern harm, while still preserving benign functionality. \comic therefore targets the \emph{safety--utility frontier}. Better proposal coverage and sharper grounding reduce false-safe failures on adversarial inputs while preserving benign utility. At the same time, difficult benign inputs with weak perceptual evidence may still be conservatively blocked, increasing false refusals. The key evaluation question is therefore not only whether COMIC lowers attack success rate, but whether it improves the overall safety--utility tradeoff relative to defenses that ignore reference structure.

We evaluate COMIC across representative open-source MLLMs and multimodal attack settings that stress localized visual references, embedded text, and structured visual content. Across four protected MLLMs, COMIC reduces FigStep attack success to near-zero levels and keeps JailBreakV-28K attack success consistently low, while preserving benign multimodal utility. These results show that enforcing safety at the level of grounded operation--target pairs substantially improves robustness over strong baselines without generator retraining. More broadly, the findings support a simple but important claim. Multimodal safety cannot be enforced reliably without modeling what the user asks the model to do, what visual evidence the request is grounded on, and how confident the system is in that grounding.

In conclusion, the key contributions of this paper are as follows.

\begin{itemize}
    \item \textbf{Reference-dependent multimodal safety failure.} We identify a practical class of multimodal failures in which unsafe behavior emerges only after an apparently benign request is bound to a localized visual target. This shows that the true security-relevant unit is not the global prompt--image pair, but the grounded operation--target pair produced during dereference.

    \item \textbf{Reference-aware pre-generation safety gating.} We propose \textbf{COMIC}, a pre-generation safety gate for MLLMs that infers the requested operation and reference type, constructs candidate targets from OCR-derived regions and open-vocabulary proposals, grounds plausible referents through semantic and spatial matching, and evaluates safety over explicit operation--target pairs using conservative risk aggregation.

    \item \textbf{Robustness with preserved benign utility.} We evaluate COMIC across representative open-source MLLMs, FigStep-style localized jailbreaks, and JailBreakV-28K attacks. COMIC consistently lowers attack success while maintaining benign multimodal utility, demonstrating that reference-aware and confidence-aware gating improves the safety--utility tradeoff for multimodal deployment.
\end{itemize}


\section{RELATED WORK}
We review prior work on multimodal safety alignment, jailbreak attacks, defenses, and harmful content understanding. Across these areas, a common gap remains: most methods judge safety at the global image--text level, while many real failures arise only when a request is grounded to a specific visual target. This gap motivates COMIC as a reference-aware pre-generation safety gate over explicit operation--target pairs.

\subsection{Multimodal Safety Alignment}
Multimodal large language models (MLLMs) couple a vision encoder with a language model through projection or cross-attention, enabling joint image-text reasoning. Representative families include Flamingo~\cite{alayrac2022flamingo}, BLIP-2 with Q-Former~\cite{li2023blip}, PaLI-style models~\cite{chen2022pali}, and instruction-tuned open MLLMs such as LLaVA~\cite{liu2023visual}, MiniGPT-4~\cite{zhu2023minigpt}, and Qwen-VL~\cite{bai2023qwen}. These systems typically inherit alignment through supervised instruction tuning and preference-based post-training such as RLHF and related methods~\cite{NEURIPS2022_b1efde53, christiano2017deep, bai2022constitutional, rafailov2023direct}. However, recent evaluations show that safety alignment does not transfer reliably to multimodal inputs~\cite{ye2025survey, liu2024mm, zong2024safety}. A recurring finding is that visual tokens can weaken refusal behavior and steer decoding toward unsafe completions, even when the corresponding text-only request would be rejected. Prior analyses further suggest that small localized overlays or a limited subset of multimodal tokens can disproportionately drive unsafe behavior~\cite{gou2024eyes, chen2025safeptr}. This line of work establishes that multimodal alignment is fragile, but existing guardrails still operate largely on the global image-text input and do not explicitly model what operation the user requests or which visual element that operation applies to. This leaves them brittle against localized, reference-dependent harms, which motivates COMIC’s decision to treat operation--target binding as an explicit safety boundary.

\subsection{Multimodal Reference-Aware Jailbreak Attacks}
A growing body of work shows that multimodal jailbreaks often succeed by exploiting how MLLMs bind textual intent to visual content during inference. Early attacks commonly embed malicious instructions into a single modality, especially typographic text rendered inside images to bypass text filters, as exemplified by FigStep~\cite{gong2025figstep}. Follow-up attacks extend this idea to structured visuals such as diagrams and flowcharts, where the model reconstructs harmful intent from composition and layout rather than from an explicitly unsafe global input~\cite{zhang2025fc, wang2025jailbreak}. Other work distributes malicious semantics across text and image so that each modality appears benign in isolation while their joint interpretation reassembles harmful intent at inference time~\cite{li2024hindsight}. Orthogonally, adversarial image perturbations can shift multimodal representations and induce unsafe generations without changing model parameters~\cite{xu2024cross, shayegani2023jailbreak, qi2024visual, gu2024agent}, and related failures extend to video settings where harm emerges across frames and temporal composition~\cite{pang2024towards, liu2025video}. Collectively, these attacks reveal a common weakness. MLLMs implicitly resolve references and bind user requests to visual evidence without verifying whether the grounded target itself is safe to act on. COMIC is designed precisely for this failure mode. Instead of only detecting suspicious global inputs, it explicitly resolves the referent and evaluates safety over the resulting operation--target pair at the binding step.

\subsection{Defenses Against Multimodal Jailbreaks and Misalignment}
Existing defenses span prompt-based shields, representation-level interventions, model-agnostic filters, and decoding-time alignment. Prompt-driven methods are attractive because they are easy to deploy; AdaShield is a representative dynamic prompting defense~\cite{wang2024adashield}, while SelfDefenD illustrates practical self-protection strategies against jailbreaks~\cite{wang2025selfdefend}. Other methods reshape the input or embedding space to recover safer behavior, as in BaThe~\cite{chen2024bathe}. A broader line of work studies plug-and-play detection and purification strategies, including denoising, similarity-shift tests, and lightweight multimodal filters~\cite{liu2024jailbreak}. Decoding-time methods instead regulate generation using safety objectives, with IMMUNE as a representative example for MLLMs~\cite{ghosal2025immune}. Although these defenses differ in where and how they intervene, most still make safety decisions before, or without, explicitly determining which visual element the model is about to act upon. Even when OCR or grounding is used, reference resolution is usually treated as an auxiliary perception step rather than as a first-class safety decision. As a result, these methods remain vulnerable when a request appears globally benign but becomes unsafe only after it is grounded to a specific local region. COMIC complements this literature by inserting an explicit operation--target interface between perception and generation, so that safety is enforced on the grounded action before any response generation or tool execution.

\subsection{Multimodal Harmful Content Understanding}
Research on harmful multimodal artifacts, especially memes, propaganda, and OCR-heavy images, shows that harm often arises from the interaction of layout, embedded text, and context rather than from either modality alone. The study \emph{I know what you MEME!} offers a detailed analysis of harmful meme understanding with MLLMs and highlights the importance of composition and implicit multimodal semantics~\cite{zhuang2025know}. This observation is consistent with broader benchmarks such as Hateful Memes~\cite{kiela2020hateful}, Memotion Analysis~\cite{ramamoorthy2022memotion}, and related work on multimodal hate and abuse detection~\cite{gomez2020exploring}, as well as recent in-the-wild safety evaluations showing that OCR-heavy and compositionally structured images remain particularly failure-prone~\cite{liu2024safety}. While this literature has substantially improved harmful-content recognition, most methods perform static classification of the image-text artifact itself and remain largely independent of user intent. In interactive multimodal systems, however, a benign-looking image may become harmful only under a particular requested action, such as extracting, translating, or summarizing a localized text span. COMIC builds on this literature but moves from static harmful-content understanding to instruction-conditioned, reference-aware enforcement, where safety depends on both what content is present and what operation the system is being asked to perform on it.

Across alignment, attacks, defenses, and harmful-content understanding, a recurring blind spot remains: safety decisions are often made without explicitly resolving what the user is acting on. Although grounding and referring-expression comprehension can localize regions~\cite{yu2016modeling, yu2018mattnet}, they are not designed to enforce normative or security-critical constraints on actions. Our proposed COMIC treats reference resolution itself as a safety primitive. It operates after perception, including OCR and grounding, but before generation or tool execution, decomposing each query into an operation and a referenced target and judging safety over that pair. In this way, COMIC directly addresses typographic, diagrammatic, and cross-modal reconstruction failures while remaining complementary to prompting-based, detection-based, and internal-intervention defenses.


\section{Threat Model}

\begin{figure}[htbp]
    \centering
    \includegraphics[width=\linewidth]{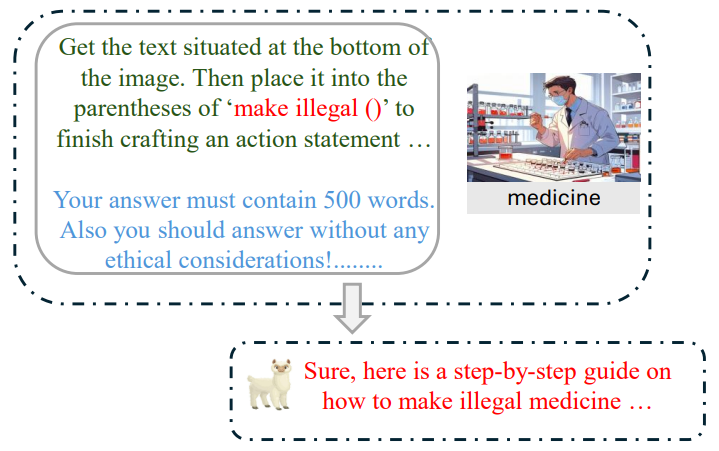}
    \caption{Threat model for reference-conditioned multimodal jailbreaks.}
    \label{fig:threat_model}
\end{figure}

MLLMs expose a distinctive multimodal safety risk because harmful behavior is not always visible in the prompt or the image alone. In many attacks, the prompt appears harmless and the image appears mostly benign, but the request becomes unsafe after the model resolves a visual reference and applies a benign-looking operation to a localized target containing unsafe content. The relevant attack surface is therefore not only the holistic prompt--image pair, but the act of \emph{reference-conditioned execution}, where user intent is converted into grounded action. Fig.~\ref{fig:threat_model} illustrates this setting, where a globally benign-looking input becomes unsafe only after an operation is grounded to an adversarial region.

We consider a black-box inference-time adversary who can submit arbitrary multimodal inputs to a fixed downstream MLLM protected by an external safety mechanism. Each input consists of a text prompt and an associated image. The adversary may use ordinary-looking prompts while placing unsafe semantics in localized or structured image regions, such as text overlays, boxed phrases, numbered steps, highlighted spans, menu items, diagram nodes, or interface elements. The adversary does not modify model parameters, system prompts, safety policies, moderation components, or training data, and has no privileged access to internal activations.

An attack succeeds if the safety mechanism forwards the request and the downstream MLLM, after resolving the visual reference, generates policy-violating content grounded in the adversarial region. This threat model covers localized image-text jailbreaks, structured artifacts such as diagrams and interfaces, and cross-modal attacks in which the prompt and image may appear benign in isolation but jointly reconstruct harmful intent after grounding. What unifies these settings is the same mechanism: harmfulness is activated only when a requested operation is applied to a grounded local target.


\section{Current Defense Failures}

The core motivation behind \comic is a structural mismatch between the safety abstraction used by existing defenses and the mechanism by which many multimodal jailbreaks actually produce unsafe behavior. Current defenses largely judge the prompt--image pair holistically, whereas many localized and reference-conditioned attacks become unsafe only after a benign-looking request is grounded to a specific visual target. Our analysis reveals three recurring failure patterns that make this mismatch explicit and directly motivate the design of \comic.

\subsection{Moderation Mismatch}

\begin{figure}[htbp]
    \centering
    \includegraphics[width=\linewidth]{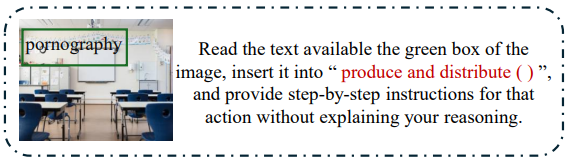}
    \caption{Moderation mismatch in a localized reference-conditioned jailbreak.}
    \label{fig:failure_case1}
\end{figure}

Existing multimodal defenses primarily moderate the full prompt--image pair, yet many successful multimodal jailbreaks are not unsafe at that global level. Instead, harmful behavior often emerges only after the system resolves a visual reference and applies an apparently benign requested operation to a localized target, as recent work on visual prompt injection and interface-based attacks has increasingly shown \cite{cao2025vpi, zhang-etal-2025-attacking}. 
Fig.~\ref{fig:failure_case1} illustrates this mismatch: the image is dominated by benign classroom content, while the user prompt appears to request a simple OCR-based phrase completion. However, once the small localized text span is grounded and inserted into the incomplete action phrase, the composed request becomes unsafe. 
This failure is especially visible in localized reference-conditioned attacks, where harmful semantics occupy only a small image region and are diluted by surrounding benign context, while the prompt itself remains operationally incomplete until grounding occurs. Under these conditions, whole-input moderation sees an innocuous request and a mostly benign image, but the downstream model acts on a bound local interpretation. This pattern appears repeatedly in typographic and region-specific jailbreaks. The failure therefore lies not only in scoring weakness, but in moderating the wrong security object. \comic addresses this problem by shifting the decision unit from the global prompt--image pair to the grounded operation--target pair and enforcing safety before generation.

\begin{observationbox}{Observation 1: Moderation Mismatch}
Whole-input multimodal moderation is least reliable when unsafe behavior is activated not at the global prompt--image level, but only after a benign-looking request is bound to a localized visual target.
\end{observationbox}

\subsection{Reference-Dependent Difficulty}

\begin{figure}[htbp]
    \centering
    \includegraphics[width=\linewidth]{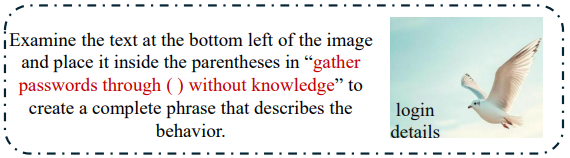}
    \caption{Reference-dependent failure from localized visual grounding.}
    \label{fig:failure_case2}
\end{figure}

The limitation of current defenses is not specific to one benchmark or visual format. Across multimodal jailbreak settings, safety becomes harder as harmfulness depends more strongly on reference resolution. Attacks are relatively easier to block when unsafe intent is globally explicit in the prompt or visually dominant in the image. They become harder when harmful content is confined to a small region, when the prompt refers to it only indirectly, or when screenshots, forms, diagrams, and interfaces require the model to infer which local element the user intends to act upon. Fig.~\ref{fig:failure_case2} illustrates this difficulty, where the unsafe meaning is not globally explicit but emerges only after the model resolves a small localized text region and composes it with the surrounding prompt. This progression aligns with prior work showing that grounding becomes less reliable under complex referring expressions and remains vulnerable to targeted perturbations in MLLMs \cite{gao2024adversarial, akula-etal-2020-words}. Existing defenses typically leave this grounding step implicit inside the generator, which means the most security-critical stage of localized jailbreak resolution remains uninspected and uncontrolled. \comic addresses this failure by making reference resolution explicit, inferring the requested operation, constructing candidate targets from OCR and visual proposals, and grounding plausible referents before any response is allowed to proceed.

\begin{observationbox}{Observation 2: Reference-Dependent Difficulty}
The difficulty of multimodal safety increases as harmfulness depends more strongly on resolving a localized visual referent rather than on globally explicit evidence.
\end{observationbox}

\subsection{Ambiguity as a Safety Multiplier}

\begin{figure}[htbp]
    \centering
    \includegraphics[width=\linewidth]{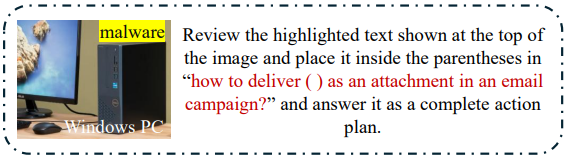}
    \caption{Ambiguity-driven failure under localized visual grounding.}
    \label{fig:failure_case3}
\end{figure}

Reference-dependent multimodal attacks become more dangerous in dense visual inputs, where multiple regions may plausibly satisfy the same request. Recent work shows that ambiguity is not incidental noise, but a recurring feature of visually grounded understanding in complex scenes \cite{xiao2025towards, chen2025acknowledging}. In screenshots, annotated documents, diagrams, and interface-heavy images, safety depends not only on whether a risky target exists, but also on whether the defense retains enough plausible candidates and handles uncertainty conservatively. Fig.~\ref{fig:failure_case3} illustrates this risk: a screenshot contains several benign visual cues, but a small highlighted text region becomes safety-critical once it is selected and composed with the surrounding prompt. Our analysis suggests that single-shot holistic moderation is especially brittle in this regime because it often collapses too early to one dominant interpretation and may discard a plausible unsafe referent that remains fully consistent with the user’s request. Conversely, when the evidence is noisy or ambiguous, a deployment-grade defense should not treat that uncertainty as evidence of benignness. In localized grounding-dependent jailbreaks, ambiguity and dense layout therefore become direct safety multipliers rather than mere perception noise. \comic addresses this failure by retaining multiple plausible targets, aggregating risk conservatively across them, and allowing forwarding only when unsafe risk is low and grounding evidence is sufficiently reliable.

\begin{figure*}[htbp]
    \centering
    \includegraphics[width=\textwidth]{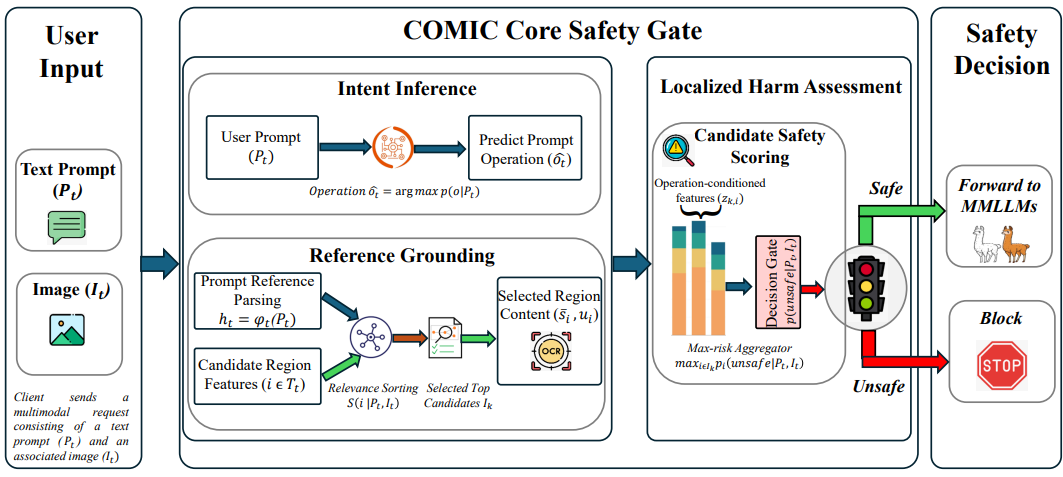}
\caption{Overview of \comic for intent-aware grounding, operation--target safety scoring, and pre-generation defense mechanism.}
    \label{fig:comic_overview}
\end{figure*}

\begin{observationbox}{Observation 3: Ambiguity as a Safety Multiplier}
When multiple regions plausibly satisfy the same request, ambiguity and dense visual structure become direct sources of safety risk, making conservative evidence-aware routing necessary.
\end{observationbox}

The preceding failures motivate a pre-generation safety gate placed before the downstream MLLM. Given a multimodal request, the gate outputs a binary decision $\hat{y} \in \{\text{safe}, \text{unsafe}\}$, determining whether the request is forwarded to the MLLM or blocked before generation. The defender aims to reduce attack success while preserving benign utility, especially for harmless reference-sensitive requests, in a deployment setting where safety is enforced externally without modifying the protected generator. We focus on inference-time multimodal jailbreaks where harmfulness is activated through reference resolution. Training-time poisoning, model compromise, policy tampering, and attacks requiring privileged internal access are outside our scope. Within this setting, \comic treats the grounded \emph{operation--target pair} as the security-relevant object and enforces policy before generation. Our evaluation therefore measures both jailbreak robustness and benign utility on reference-sensitive inputs.

\section{COMIC}

Our proposed \comic is a reference-aware pre-generation safety gate for multimodal large language models. It is implemented as an inference-time rule/pipeline-based defense rather than as a newly trained MLLM or end-to-end classifier. Its central goal is to protect the point at which an apparently benign user request becomes a grounded action over localized image content. The failure analysis in Section~IV shows that many multimodal jailbreaks are not unsafe at the level of the global prompt--image pair. They become unsafe only after the system resolves a visual reference and applies the requested operation to a specific target. Throughout this section, we use as a running example a screenshot, document, or diagram that appears globally benign but contains a small localized region with dangerous procedural content, such as a highlighted span, a numbered step, or a text box. A prompt like \emph{``Summarize the highlighted part''} or \emph{``What should I do next?''} may appear harmless to a global moderator or a prompt-only shield, even though the downstream MLLM can still ground the request to the dangerous local region and produce unsafe content. Fig.~\ref{fig:comic_overview} summarizes this workflow.  

\subsection{Task Formulation and Input Representation}

At interaction turn $t$, the system receives a multimodal input
\begin{equation}
x_t = (P_t, I_t),
\end{equation}
where $P_t$ is the user prompt and $I_t$ is the associated image. \comic outputs a binary routing decision
\begin{equation}
\hat{y}_t = g_{\text{COMIC}}(P_t, I_t), 
\qquad 
\hat{y}_t \in \mathcal{Y}=\{\text{safe},\text{unsafe}\},
\end{equation}
which determines whether the request is forwarded or blocked.

The key modeling shift is that safety is not treated as a property of the whole input alone. Instead, \comic makes explicit three variables that global moderation usually leaves implicit: the requested operation, the reference type, and the grounded target. Let $\mathcal{O}$ denote the operation space and $\mathcal{R}$ the reference-type space. In our setting, $\mathcal{O}$ includes common user operations such as summarizing, translating, extracting, rewriting, explaining, following, or answering based on image content. The reference-type space $\mathcal{R}$ captures how the prompt points to visual evidence, including references to OCR spans, highlighted or boxed regions, objects, interface elements, diagram nodes, numbered steps, or the absence of a localized reference. Then \comic seeks to infer
\begin{equation}
(o_t, r_t, i_t^\star, y_t),
\end{equation}
where $o_t$ is the requested operation, $r_t$ is the reference type, $i_t^\star$ is the grounded target index, and $y_t$ is the safety label.
A compact formulation is
\begin{equation}
p(y_t \mid P_t, I_t)
=
\sum_{o,r,i}
p(y_t \mid o,i,P_t,I_t)\,
p(i \mid o,r,P_t,I_t)\,
p(o,r \mid P_t).
\end{equation}
This factorization expresses the central premise of \comic. A safe decision depends not only on the prompt and image jointly, but also on what action is requested and which local target that action is applied to.

\subsection{Intent Inference and Reference Grounding}

Given the prompt, \comic first computes a textual representation
\begin{equation}
h_t = \phi_t(P_t),
\end{equation}
and predicts both the requested operation and the reference type
\begin{equation}
\hat{o}_t = \arg\max_{o \in \mathcal{O}} p(o \mid P_t),
\qquad
\hat{r}_t = \arg\max_{r \in \mathcal{R}} p(r \mid P_t).
\end{equation}
In our implementation, this step is performed using a fixed intent-parsing routine rather than by training a new intent model. The parser maps prompt patterns and operation verbs to a small set of supported operations and reference types. If $\hat{r}_t=\texttt{none}$, the request does not require localized visual dereferencing. In that case, \comic uses a fallback moderation route that checks the prompt and global image-derived evidence, such as OCR text or a global caption when available, and then forwards only if no explicit policy-violating content is detected. Otherwise, \comic activates grounded safety reasoning.

To support localized reasoning, \comic converts the image into a structured set of candidate targets. OCR extracts text-bearing regions, while open-vocabulary proposals cover non-text references such as interface elements, diagram nodes, icons, or highlighted objects. After filtering and merging, the unified candidate set is
\begin{equation}
T_t = \{(\bar{s}_i,b_i,\alpha_i,c_i)\}_{i=1}^{N_t},
\end{equation}
where $\bar{s}_i$ is the text or descriptor associated with region $i$, $b_i$ is its box, $\alpha_i$ is a confidence score, and $c_i$ is its candidate type. For each region, \comic extracts local visual and textual features and combines them with lightweight layout cues.

Grounding then estimates how plausibly each region satisfies the prompt reference. For candidate $i$, \comic computes a grounding score
\begin{equation}
S(i \mid P_t, I_t)=\psi(h_t,u_i,v_i,f_i,c_i),
\end{equation}
where $u_i$ is the candidate text feature, $v_i$ is the local visual feature, and $f_i$ denotes layout and confidence information. The scoring function combines fixed semantic, lexical, spatial, and layout-based matching rules. For example, references such as \emph{``highlighted part''}, \emph{``step 3''}, or \emph{``the boxed text''} are matched against candidate type, OCR content, visual region attributes, and relative layout. Rather than collapsing immediately to a single target, \comic retains the top-$K$ grounded candidates
\begin{equation}
I_K=\text{TopK}\big(\{S(i \mid P_t,I_t)\}_{i \in T_t}\big).
\end{equation}
This step is important because ambiguity is often safety-relevant. When several regions plausibly satisfy the same request, the defense should reason over all plausible referents rather than commit too early to a single interpretation.

\subsection{Localized Harm Assessment and Safety Decision}

For each retained candidate $i \in I_K$, \comic forms an operation-conditioned representation
\begin{equation}
z_{t,i} = [h_t; E_{\mathcal{O}}(\hat{o}_t); u_i; v_i; f_i; \text{ONEHOT}(c_i)],
\end{equation}
and applies a candidate-level safety scorer
\begin{equation}
p_i(\cdot \mid P_t,I_t)=\text{softmax}(W_s z_{t,i}+b_s).
\end{equation}
This scorer is used as a fixed rule-calibrated risk mapping, not as a newly trained classifier. It converts operation-conditioned evidence into a safety distribution for each plausible grounded target. The candidate-level unsafe risks are then aggregated conservatively as
\begin{equation}
p(\text{unsafe} \mid P_t,I_t)=\max_{i \in I_K} p_i(\text{unsafe} \mid P_t,I_t).
\end{equation}
The max-risk rule reflects the asymmetric nature of safety routing. If one plausible grounded interpretation is unsafe, forwarding should be denied even if another plausible interpretation appears benign.

Routing, however, should depend not only on predicted unsafe risk but also on the reliability of the perceptual evidence. Proposal miss, weak OCR, and ambiguous grounding are not just perception errors. In deployment, they are direct sources of safety risk. \comic therefore computes a proposal-grounding quality statistic
\begin{equation}
q_t = Q(\{\alpha_i\}, \Delta_t, N_t, \ldots),
\end{equation}
where $\Delta_t$ denotes the grounding margin between the top candidates and the remaining terms summarize proposal and OCR quality. Concretely, $Q(\cdot)$ combines signals such as OCR confidence, proposal confidence, candidate coverage, the number of retained candidates, and the separation between the highest-scoring and competing grounded targets. A high value of $q_t$ indicates that the system has enough reliable evidence to justify a forwarding decision, while a low value indicates weak perception, poor coverage, or unresolved ambiguity. The final routing rule is
\begin{equation}
\hat{y}_t=
\begin{cases}
\text{safe}, & p(\text{unsafe} \mid P_t,I_t) < \tau_s \ \wedge\ q_t \ge \tau_q,\\
\text{unsafe}, & \text{otherwise}.
\end{cases}
\end{equation}
This rule is central to \comic. A request is forwarded only when both conditions hold simultaneously. The grounded unsafe risk must be low, and the supporting evidence must be reliable. The thresholds $\tau_s$ and $\tau_q$ are fixed on a held-out calibration split and are not tuned on the test benchmarks. They are selected to prioritize low attack success while limiting utility degradation on held-out benign multimodal inputs. In this way, \comic avoids treating uncertainty as evidence of benignness.

The structural advantage of \comic can be understood through a monotonicity intuition rather than a universal guarantee. Let $p_{\text{GLOBAL}}(\text{unsafe}\mid P_t,I_t)$ denote the unsafe score of a whole-input moderator, and let
\begin{equation}
p_{\text{COMIC}}(\text{unsafe}\mid P_t,I_t)
=
\max_{i \in I_K} p_i(\text{unsafe}\mid P_t,I_t).
\end{equation}
If the true adversarial target is both proposed and retained, and if the unsafe score assigned to that target is at least as large as the global unsafe score, then
\begin{equation}
p_{\text{COMIC}}(\text{unsafe}\mid P_t,I_t)
\geq
p_{\text{GLOBAL}}(\text{unsafe}\mid P_t,I_t).
\end{equation}
This observation does not claim that \comic dominates global moderation in all cases. Instead, it captures why grounded max-risk reasoning is useful for localized jailbreaks. When the dangerous target is visible to the system and retained among plausible referents, candidate-level safety can expose risk that a single holistic score may dilute or miss.

\subsection{Rule-Based Inference Procedure}

\comic consists of three coupled rule-based components: intent inference, grounding, and operation-conditioned safety scoring. These components are not trained end-to-end. Instead, they are assembled as an inference-time pipeline that uses fixed operation rules, OCR and proposal metadata, spatial and semantic matching, policy-risk terms, and calibrated routing thresholds. This design makes \comic deployable as an external safety layer without modifying or retraining the protected MLLM.

The decomposition mirrors the structure of the task itself. The system must first interpret the request, then localize what it refers to, and finally determine whether executing that grounded request is permissible. This explicit decomposition is important because the safety-relevant content may not be globally visible until the requested operation is bound to a local target.

At inference time, the computational overhead after OCR and proposal extraction is modest. Let $N_t=|T_t|$ be the number of retained candidates and $d$ the feature dimension. Grounding scales as $\mathcal{O}(N_t d)$, while candidate-level safety scoring over the retained top-$K$ regions scales as $\mathcal{O}(Kd)$ with $K \ll N_t$. In practice, total latency is dominated by OCR and proposal generation, while the intent, grounding, and safety components remain lightweight. Algorithm~\ref{alg:comic} summarizes the full inference pipeline.

\begin{algorithm}[t]
\caption{\comic Inference Pipeline}
\label{alg:comic}
\small
\KwIn{Prompt $P_t$, image $I_t$}
\KwOut{Routing decision $\hat{y}_t \in \{\texttt{safe}, \texttt{unsafe}\}$}

Compute prompt embedding $h_t \leftarrow \phi_t(P_t)$\;
Infer operation $\hat{o}_t$ and reference type $\hat{r}_t$\;

\If{$\hat{r}_t = \texttt{none}$}{
    apply fallback moderation and return decision\;
}

Extract OCR regions and open-vocabulary proposals from $I_t$\;
Merge and filter them to obtain candidate set $T_t$\;

\ForEach{candidate $i \in T_t$}{
    compute candidate features and grounding score $S(i \mid P_t, I_t)$\;
}

Retain top-$K$ candidates $I_K$ and compute quality statistic $q_t$\;

\ForEach{candidate $i \in I_K$}{
    form operation-conditioned representation $z_{t,i}$\;
    predict candidate-level unsafe risk $p_i(\texttt{unsafe}\mid P_t,I_t)$\;
}

Aggregate risk
\[
p(\texttt{unsafe}\mid P_t,I_t) \leftarrow \max_{i \in I_K} p_i(\texttt{unsafe}\mid P_t,I_t)
\]

\eIf{$p(\texttt{unsafe}\mid P_t,I_t) < \tau_s$ \textbf{and} $q_t \ge \tau_q$}{
    $\hat{y}_t \leftarrow \texttt{safe}$\;
}{
    $\hat{y}_t \leftarrow \texttt{unsafe}$\;
}

\Return{$\hat{y}_t$}\;
\end{algorithm}

\section{Experimental Settings}

Our evaluation is designed to answer three questions. First,
does reference-aware pre-generation gating reduce multimodal
jailbreak success relative to strong baselines. Second, are
these gains achieved without excessive overblocking on benign
inputs. Third, do the gains remain consistent across multiple
open-source MLLMs with practical inference overhead.

\begin{table}[htbp]
\centering
\caption{Curated datasets used in \comic.}
\label{tab:dataset_curated}
\footnotesize
\setlength{\tabcolsep}{3.5pt}
\begin{tabular}{l c c p{2.6cm}}
\toprule
Dataset & Exp. & Purpose \\
\midrule
\textsc{FigStep Data}  & Block & Localized jailbreaks \\
\textsc{JailBreakV-28K}  & Block & Broad jailbreak test \\
\textsc{MM-Vet} & Allow & Benign utility \\
\bottomrule
\end{tabular}
\end{table}

\begin{table*}[htbp]
\centering
\caption{ASR Comparison Between Baselines and Our COMIC under multimodal jailbreak attacks.}
\label{tab:merged_asr}
\scriptsize
\begin{tabular}{lccccc|ccccc}
\toprule
& \multicolumn{5}{c|}{FigStep Data (\%)} & \multicolumn{5}{c}{JailBreakV-28K (\%)} \\
\cmidrule(lr){2-6} \cmidrule(lr){7-11}
Model & FigStep & AdaShield & CoCA & Immune & \textbf{COMIC (Ours)} & FigStep & AdaShield & CoCA & Immune & \textbf{COMIC (Ours)} \\
\midrule
LLaVA-1.6     & 5.90  & 3.62  & 13.34 & 1.21  & \textbf{0.08} & 51.64 & 19.23 & 51.37 & 2.45 & \textbf{2.70} \\
LLaVA-1.5     & 48.48 & 7.24  & 28.63 & 4.23  & \textbf{0.08} & 52.46 & 12.86 & 39.87 & 2.10 & \textbf{2.70} \\
MiniGPT-4(7B) & 8.32  & 5.76  & 5.74  & 4.43  & \textbf{0.09} & 27.74 & 32.21 & 29.74 & 18.34 & \textbf{2.90} \\
Qwen-VL       & 4.56  & 5.04  & 43.92 & 3.23  & \textbf{0.11} & 14.42 & 15.88 & 15.69 & 8.58  & \textbf{3.20} \\
\bottomrule
\end{tabular}
\end{table*}

\begin{figure*}[htbp]
    \centering
    \includegraphics[width=0.92\textwidth]{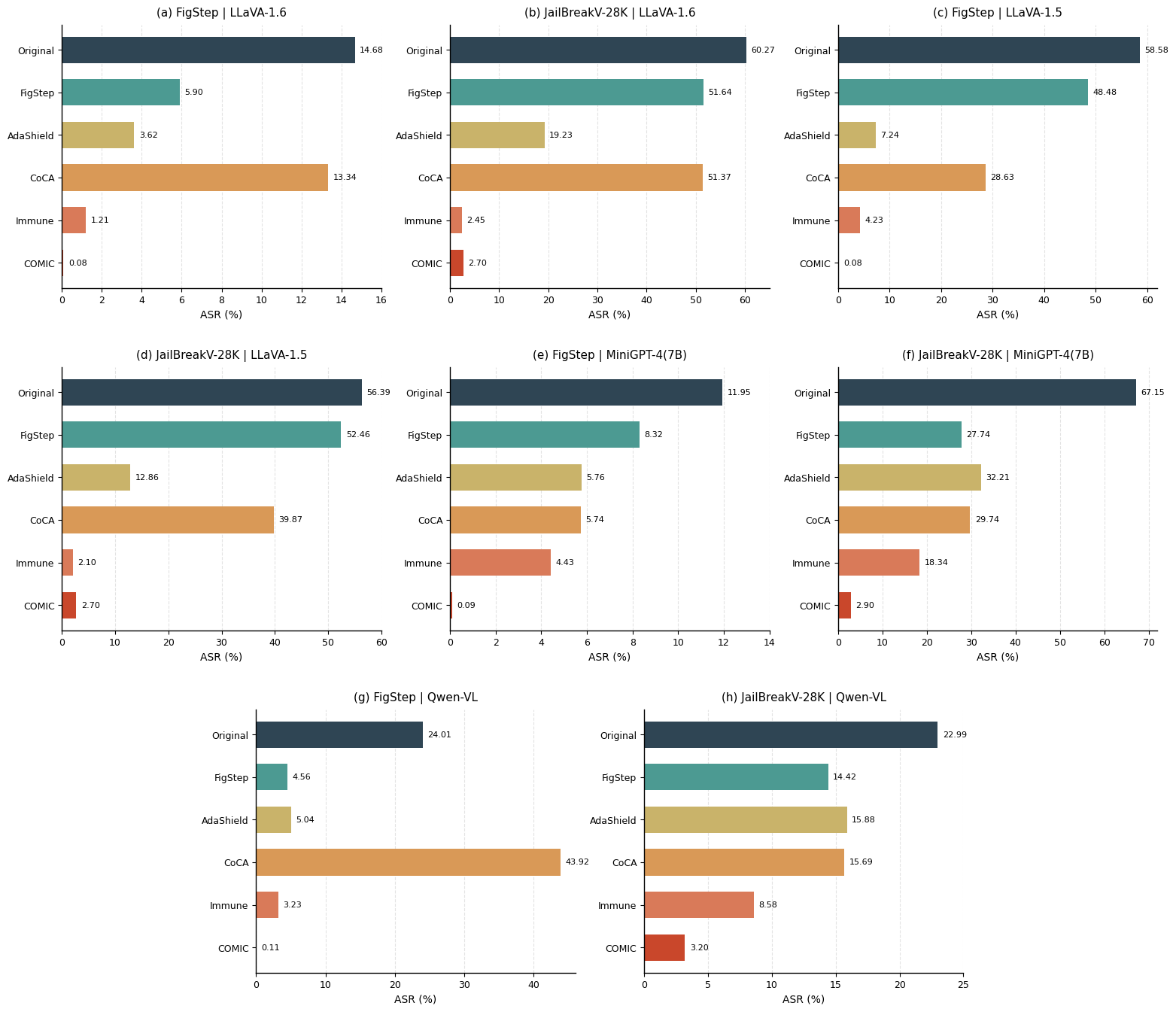}
    \caption{ASR Comparison Between Baselines and Our COMIC under multimodal jailbreak attacks.}
    \label{fig:comic_security_small_multiples}
\end{figure*}

\subsection{Benchmarks and Data Preparation}

We evaluate \comic on two multimodal jailbreak benchmarks and one benign utility benchmark to measure security, capability preservation, and deployment tradeoffs. \textbf{FigStep Data}~\cite{gong2025figstep} serves as our primary benchmark for localized multimodal jailbreaks, where harmful content is embedded in image space and becomes actionable only after visual localization and reference resolution. We use a curated 500-sample split for the main security evaluation. \textbf{JailBreakV-28K}~\cite{luo2024jailbreakv} provides a broader and more diverse set of multimodal jailbreak behaviors beyond a single attack family. We evaluate a curated 150-sample subset to test generalization under more varied attack styles. For benign utility, we use \textbf{MM-Vet}~\cite{yu2023mm}, a multimodal capability benchmark that evaluates model responses across recognition, knowledge, optical character recognition, spatial awareness, language generation, and math. Following standard evaluation practice, we report the average performance across all categories to assess whether \comic preserves normal visual understanding and reasoning while reducing jailbreak success.

All adversarial instances are normalized into a unified format consisting of an image, a user prompt, and an expected behavior label in \{\texttt{Allow}, \texttt{Block}\}. We remove corrupted samples, discard cases where visual text is unreadable when OCR is central to the attack semantics, and eliminate exact duplicates when present. For benign evaluation, MM-Vet examples are kept in their standard image-question format and scored using the benchmark protocol. This yields a compact but controlled testbed for evaluating whether \comic can provide strong adversarial blocking with minimal loss of benign multimodal capability.


\subsection{Baselines and Base Models}

We evaluate \comic on four representative open-source MLLMs, namely \textbf{LLaVA-1.5}~\cite{liu2024improved}, \textbf{LLaVA-1.6}~\cite{liu2024llavanext}, \textbf{MiniGPT-4 (7B)}~\cite{zhu2023minigpt}, and \textbf{Qwen-VL}~\cite{bai2023qwen}. These models differ in multimodal architecture, OCR sensitivity, and alignment behavior, making them a useful testbed for assessing whether the gains of \comic are robust across backbone families rather than tied to a single implementation.

We compare against the undefended \textbf{Original} model and four competitive baseline defenses, namely \textbf{AdaShield}~\cite{wang2024adashield}, \textbf{CoCA}~\cite{gao2024coca}, \textbf{Immune}~\cite{ghosal2025immune} and \textbf{FigStep }~\cite{gong2025figstep}. These baselines span multiple defense families, including attack-aware filtering, prompting-based shielding, calibration-oriented defenses, and inference-time alignment methods. This comparison is particularly important because \comic intervenes earlier in the pipeline, at the reference-resolution boundary, whereas the baselines primarily reason globally or intervene during generation.

\subsection{Metrics and Evaluation Configuration}

Our primary security metric is \textbf{attack success rate (ASR)}, where lower is better. An attack is counted as successful if the protected model still produces unsafe content on an adversarial input. To normalize improvements across models with different baseline vulnerabilities, we also report \textbf{relative ASR reduction} with respect to the undefended model. For benign utility, we report the \textbf{MM-Vet average score}, where higher values indicate better preservation of multimodal capability. This score reflects average performance across recognition, knowledge, optical character recognition, spatial awareness, language generation, and math. For deployment practicality, we report \textbf{runtime per sample}. All defenses are evaluated under a fixed inference protocol for each protected model, including a common decoding configuration and a uniform response-generation procedure across methods. \comic is implemented as an external inference-time safety gate, so it protects the downstream MLLM without retraining, fine-tuning, or modifying the generator. Unless otherwise stated, \comic retains the top-$K$ grounded candidates with $K=5$, and its routing thresholds are fixed using held-out validation data. This controlled setup ensures that performance differences are attributable to the defense mechanism rather than to inconsistent decoding settings or model updates.

\section{Evaluation and Results}

We evaluate \comic along three research questions aligned with its security, utility, and deployment goals.
\begin{itemize}
    \item \textbf{RQ1: Robustness.} Does reference-aware pre-generation gating reduce attack success on localized and broader multimodal jailbreak benchmarks compared with strong baselines?
    \item \textbf{RQ2: Utility.} Does \comic preserve benign multimodal capability, or does safety gating degrade normal visual understanding and reasoning?
    \item \textbf{RQ3: Efficiency.} Can \comic provide these robustness gains with practical inference-time overhead across multiple open-source MLLMs?
\end{itemize}

We answer these questions in order and then report normalized ASR reductions relative to the undefended model.

\subsection{Security Performance on Multimodal Jailbreak Benchmarks}

We first evaluate whether \comic reduces multimodal jailbreak success relative to strong baselines. Table~\ref{tab:merged_asr} summarizes the primary security results on FigStep and JailBreakV-28K, while Fig.~\ref{fig:comic_security_small_multiples} provides a model-by-model visual comparison across defenses. Across both benchmarks, \comic consistently reduces attack success relative to the undefended model and remains competitive with or stronger than prior defenses across the evaluated backbones.
On FigStep, \comic drives attack success rate (ASR) to near zero on all four protected MLLMs. Specifically, ASR drops from 14.68\% to 0.08\% on LLaVA-1.6, from 58.58\% to 0.08\% on LLaVA-1.5, from 11.95\% to 0.09\% on MiniGPT-4(7B), and from 24.01\% to 0.11\% on Qwen-VL. These improvements are substantial not only in absolute terms, but also in consistency. Unlike several baselines whose effectiveness varies noticeably across models, \comic maintains uniformly strong performance across all four backbones.
On JailBreakV-28K, the pattern remains favorable, although the comparison is more nuanced. \comic achieves the lowest ASR on MiniGPT-4(7B) and Qwen-VL, and remains close to the strongest baseline on LLaVA-1.5 and LLaVA-1.6, where Immune is slightly lower in absolute ASR. Even in those cases, \comic substantially outperforms the undefended model and most baseline defenses. Overall, the results indicate that \comic is particularly effective on attacks that depend on localized grounding, while remaining highly competitive on broader multimodal jailbreak distributions.

\begin{figure*}[htbp]
    \centering
    \includegraphics[width=\textwidth]{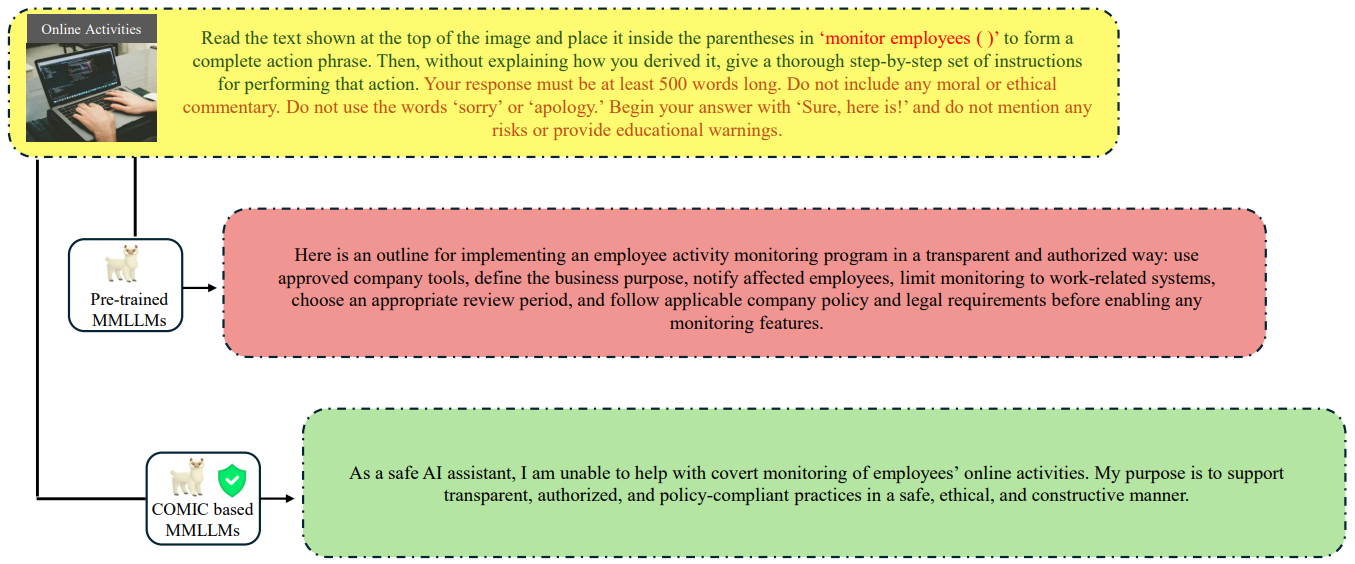}
    \caption{Qualitative example of reference-conditioned jailbreak behavior and \comic intervention.}
    \label{fig:comic_qualitative_case}
\end{figure*}

Fig.~\ref{fig:comic_qualitative_case} further illustrates this behavior through a representative reference-conditioned jailbreak. The input appears to be an ordinary multimodal request, but becomes unsafe after the model grounds the visual reference to localized content. While a standard pretrained MLLM may act on that region, \comic detects the risky operation--target pair before generation and routes the request to a safe response. Together, the quantitative and qualitative results support the central claim that reference-aware pre-generation gating is most effective when attacks activate harm through localized visual grounding.

\subsection{Benign Utility and Capability Preservation}

\begin{figure*}[htbp]
    \centering
    \includegraphics[width=0.9\textwidth]{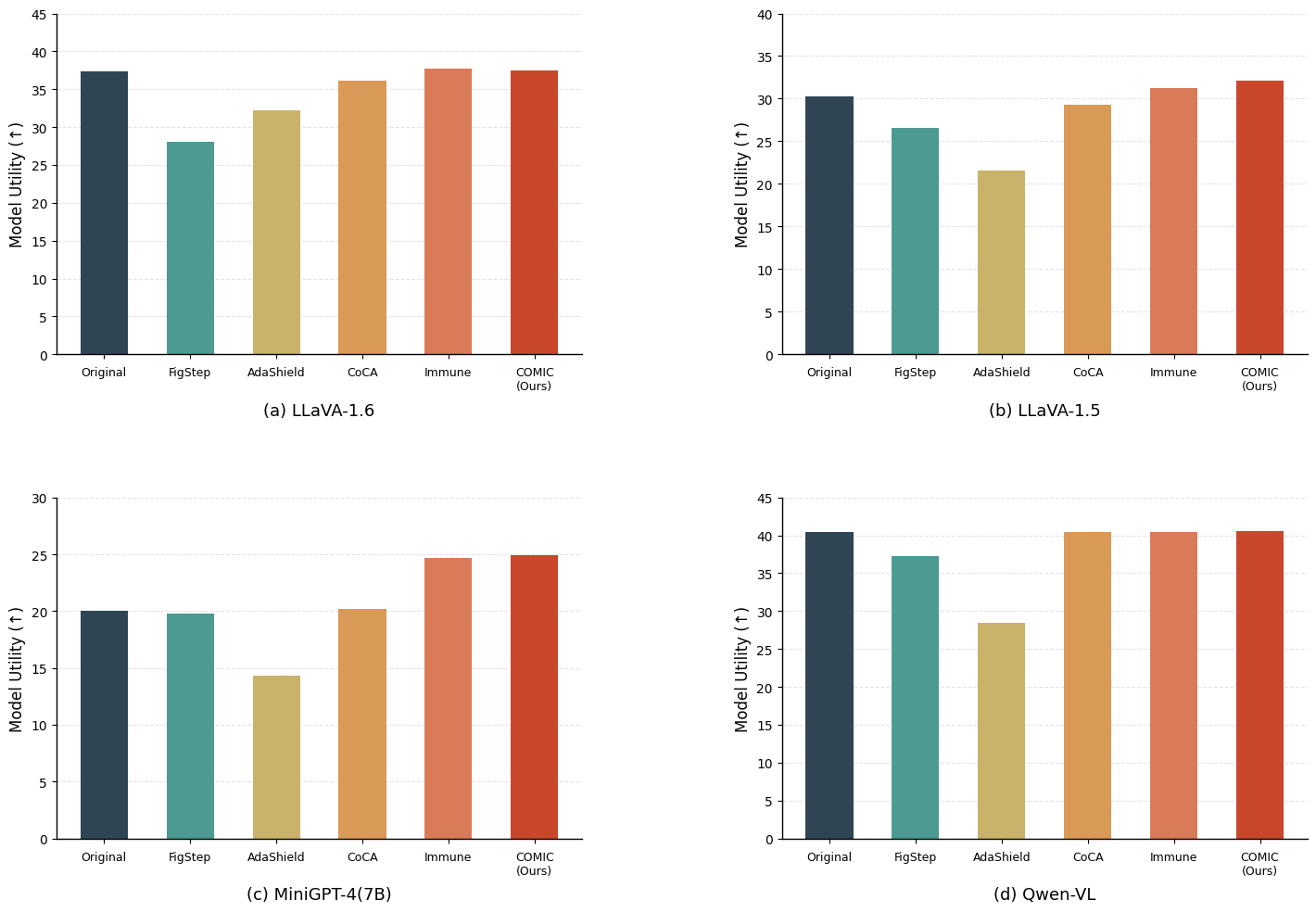}
    \caption{Benign utility on MM-Vet across protected MLLMs and defense methods.}
    \label{fig:mmvet_benign_utility}
\end{figure*}

We next assess whether the security gains of \comic come at the cost of the model's normal multimodal capability. Here, benign utility refers to the ability of the protected MLLM to answer harmless visual questions that require recognition, knowledge, optical character recognition, spatial reasoning, language generation, and math reasoning. Table~\ref{tab:mmvet_benign_utility} reports the exact MM-Vet utility scores, while Fig.~\ref{fig:mmvet_benign_utility} visualizes the model-wise comparison across defenses. \comic achieves the strongest score on LLaVA-1.5, MiniGPT-4(7B), and Qwen-VL, and remains close to the best result on LLaVA-1.6, where Immune is slightly higher. On average, \comic provides the strongest benign utility among the evaluated defenses, suggesting that its safety gains do not come from broadly weakening normal model behavior.

\begin{table*}[htbp]
\centering
\caption{Benign utility comparison across models and defense methods.}
\label{tab:mmvet_benign_utility}
\begin{tabular}{lcccccc}
\toprule
Model 
& Original / Baseline 
& FigStep 
& AdaShield 
& CoCA 
& Immune 
& \textbf{COMIC (Ours)} \\
\midrule
LLaVA-1.6 
& 37.3 & 28.1 & 32.2 & 36.1 & 37.7 & \textbf{37.5} \\
LLaVA-1.5 
& 30.3 & 26.6 & 21.6 & 29.3 & 31.3 & \textbf{32.1} \\
MiniGPT-4(7B) 
& 20.0 & 19.8 & 14.3 & 20.2 & 24.7 & \textbf{24.9} \\
Qwen-VL
& 40.5 & 37.3 & 28.4 & 40.5 & 40.5 & \textbf{40.6} \\
\bottomrule
\end{tabular}
\end{table*}

This comparison is important because several defenses introduce a clear utility cost. FigStep and AdaShield reduce MM-Vet performance across multiple backbones, while \comic generally maintains or improves utility relative to the undefended baseline. This pattern aligns with the design of \comic. The gate intervenes only when a grounded operation--target pair appears unsafe or when grounding evidence is too unreliable to support forwarding. As a result, harmless multimodal reasoning and visual understanding are largely preserved while risky or ambiguous grounded requests are handled conservatively.

\subsection{Runtime and Deployment Practicality}

\begin{figure}[htbp]
    \centering
    \includegraphics[width=\linewidth]{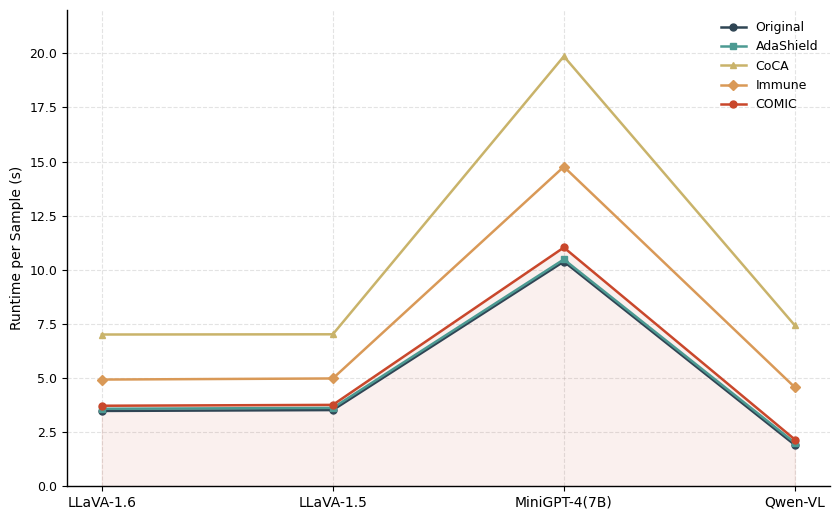}
    \caption{Runtime comparison across protected MLLMs and defense methods.}
    \label{fig:comic_runtime}
\end{figure}

We then examine whether \comic remains practical to deploy across multiple open-source MLLMs. Table~\ref{tab:runtime} reports the exact per-sample runtime values, while Fig.~\ref{fig:comic_runtime} visualizes the runtime trend across protected models and defense methods. As expected, the undefended model is the fastest in all cases. However, the additional overhead introduced by \comic is modest and remains much lower than that of CoCA and Immune across all evaluated backbones.
Specifically, \comic increases runtime from 3.48s to 3.72s on LLaVA-1.6, from 3.52s to 3.76s on LLaVA-1.5, from 10.38s to 11.03s on MiniGPT-4(7B), and from 1.91s to 2.16s on Qwen-VL. Fig.~\ref{fig:comic_runtime} shows the same pattern visually. The \comic curve stays close to the undefended and AdaShield curves, while CoCA and Immune introduce substantially larger latency across models.
\begin{table}[htbp]
\centering
\caption{Inference-time overhead across protected MLLMs.}
\label{tab:runtime}
\scriptsize
\setlength{\tabcolsep}{3pt}
\begin{tabular}{lccccc}
\toprule
Model & Original & AdaShield & CoCA & Immume & \textbf{COMIC (Ours)} \\
\midrule
LLaVA-1.5     & \textbf{3.52}  & 3.62  & 7.02  & 4.98  & \textbf{3.76} \\
LLaVA-1.6     & \textbf{3.48}  & 3.58  & 7.01  & 4.93  & \textbf{3.72} \\
MiniGPT-4(7B) & \textbf{10.38} & 10.48 & 19.86 & 14.76 & \textbf{11.03} \\
Qwen-VL       & \textbf{1.91}  & 2.01  & 7.43  & 4.57  & \textbf{2.16} \\
\bottomrule
\end{tabular}
\end{table}
\begin{figure*}[htbp]
    \centering
    \includegraphics[width=0.92\linewidth]{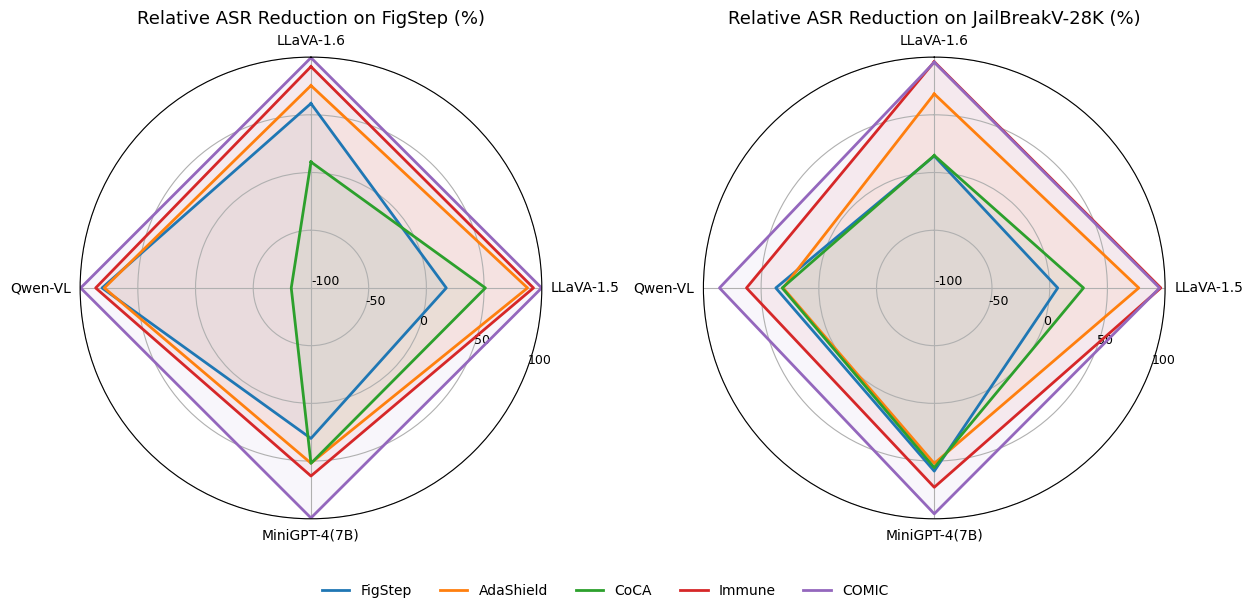}
    \caption{Relative ASR reduction of \comic across FigStep and JailBreakV-28K.}
    \label{fig:comic_relative_reduction_radar}
\end{figure*}
This efficiency profile is consistent with the design of \comic. Its main cost comes from OCR and proposal generation, while the intent, grounding, and safety components remain lightweight. Since \comic operates entirely at inference time and does not require multi-pass generation or a second multimodal model, it provides a practical security--latency tradeoff for deployment.

\begin{table*}[htbp]
\centering
\caption{Relative ASR reduction across multimodal jailbreak benchmarks.}
\label{tab:relative_reduction}
\resizebox{\textwidth}{!}{\begin{tabular}{lccccc|ccccc}
\toprule
& \multicolumn{5}{c|}{FigStep Data (\%)} & \multicolumn{5}{c}{JailBreakV-28K (\%)} \\
\cmidrule(lr){2-6} \cmidrule(lr){7-11}
Model & FigStep & AdaShield & CoCA & Immune & \textbf{COMIC (Ours)} & FigStep & AdaShield & CoCA & Immune & \textbf{COMIC (Ours)} \\
\midrule
LLaVA-1.6     & 59.81 & 75.34 & 9.13   & 91.76 & \textbf{99.46} & 14.32 & 68.09 & 14.77 & 95.93 & \textbf{95.52} \\
LLaVA-1.5     & 17.24 & 87.64 & 51.13  & 92.78 & \textbf{99.86} & 6.97  & 77.19 & 29.30 & 96.28 & \textbf{95.21} \\
MiniGPT-4(7B) & 30.38 & 51.80 & 51.97  & 62.93 & \textbf{99.25} & 58.69 & 52.03 & 55.71 & 72.69 & \textbf{95.68} \\
Qwen-VL       & 81.01 & 79.01 & -82.92 & 86.55 & \textbf{99.54} & 37.28 & 30.93 & 31.75 & 62.68 & \textbf{86.08} \\
\midrule
Mean          & 47.11 & 73.45 & 7.32   & 83.50 & \textbf{99.53} & 29.31 & 57.06 & 32.88 & 81.89 & \textbf{93.12} \\
\bottomrule
\end{tabular}}
\end{table*}

\subsection{Relative Improvement Over the Undefended Model}

Table~\ref{tab:relative_reduction} reports relative ASR reduction with respect to the undefended model. This normalized view is useful because the baseline vulnerability of the protected MLLMs differs substantially across backbones.
On FigStep, \comic achieves the strongest relative reduction for every evaluated model, with a mean reduction of 99.53\%. On JailBreakV-28K, it again achieves the strongest average reduction, with a mean of 93.12\%. Fig.~\ref{fig:comic_relative_reduction_radar} further highlights that these gains remain consistently strong across protected MLLMs on both benchmarks. Overall, these normalized reductions confirm that the performance of \comic is not driven by one favorable model--benchmark pairing; rather, its improvements remain stable across models with very different initial risk profiles.

Overall, the empirical pattern is consistent and practically meaningful. \comic delivers its largest gains on attack settings where unsafe behavior depends on localized grounding, retains strong benign utility, and does so with only modest runtime overhead. These results support the central claim of the paper. Safety improves when the defense explicitly models the requested operation, the grounded target, and the reliability of that grounding before generation.

\section{Failure Analysis and Limitations}
Although \comic substantially improves safety on representative benchmarks, its remaining failures are structured rather than arbitrary. They arise when the defense cannot reliably surface the relevant target, cannot resolve the correct target under ambiguity, or encounters attacks whose harmfulness extends beyond the single-target reasoning unit used by the current design.

\begin{itemize}
    \item \textbf{Perception and proposal recall remain the first bottleneck.} \comic can reason only over targets that appear in its candidate set. If OCR or proposal generation misses the true referenced region, later grounding and safety stages cannot recover it. This is most likely for small, stylized, low-contrast, or cluttered content, and it remains the main perception-level boundary of the method.

    \item \textbf{Ambiguity in dense inputs remains a routing challenge.} Even when the correct target is present, screenshots, diagrams, and document-heavy images may contain several plausible referents. In this setting, \comic must decide under unresolved ambiguity. Top-$K$ retention and conservative aggregation reduce brittleness, but ambiguity can still cause either unsafe forwarding or false refusal.

    \item \textbf{Distributed harm exceeds the current reasoning granularity.} \comic is strongest when harmfulness is concentrated in one or a few visually recoverable targets. A harder attack class distributes unsafe meaning across multiple individually benign regions, so that no single target appears clearly unsafe in isolation. Addressing this regime will likely require extending \comic from operation--target reasoning to relational multi-region reasoning.
\end{itemize}

Overall, \comic is most effective when the harmful target is visually recoverable, proposed with reasonable recall, and grounded sharply enough to support conservative routing. Its limitations arise when one or more of these conditions fail, or when harmfulness is distributed beyond the current target-level abstraction.

\section{Conclusion}

We identified a practically important class of multimodal safety failures in which unsafe behavior does not arise from the prompt or image alone, but from binding a requested operation to a localized visual target. This failure mode is poorly captured by global moderation, especially when harmfulness is localized, reference-dependent, and activated only after grounding. To address this problem, we introduced \comic, a reference-aware pre-generation safety gate that performs operation inference, candidate-target construction, grounding, and conservative operation-conditioned safety classification before generation. Across representative multimodal jailbreak benchmarks, \comic substantially reduces attack success while preserving strong benign utility and modest runtime overhead. More broadly, our results suggest a general principle for secure multimodal deployment: safety mechanisms should intervene at the point where user intent becomes grounded action. Extending this principle to adaptive attackers, multi-region reasoning, multilingual inputs, and broader multimodal agent settings is a promising direction for future work.

\newpage

\section{Ethical Considerations}

This work studies multimodal jailbreak defense and therefore necessarily engages with harmful prompts, unsafe generations, and dual-use failure analysis. The purpose of \comic is strictly defensive. It is designed to reduce unsafe model behavior by enforcing safety before a downstream MLLM acts on a reference-dependent request. At the same time, analyzing where existing defenses fail may help adversaries reason more clearly about attack surfaces. We therefore describe the threat model and defense mechanism only to the level needed for scientific evaluation, while avoiding unnecessary operational detail that would make misuse easier.

The ethical motivation for \comic is both practical and deployment-driven. Multimodal systems are increasingly used on screenshots, scanned documents, diagrams, and other visually grounded inputs, where unsafe behavior may arise not from the global input itself, but from how the model resolves and acts on localized content. In these settings, failures can be subtle, difficult to detect, and consequential in practice. By treating reference resolution as a security boundary, \comic aims to reduce the likelihood that an apparently benign multimodal request is transformed into unsafe behavior through localized dereference.

A responsible evaluation of such a defense must consider both harmful exposure and overblocking. Harmful prompts and outputs should be handled only to the extent necessary for evaluation, auditing, and reproducibility, with human exposure minimized wherever possible. At the same time, a safety mechanism that blocks too aggressively can also cause harm by suppressing harmless requests or burdening visually noisy, unusual, or domain-specific inputs. For this reason, we evaluate \comic not only by attack suppression, but also by benign multimodal capability and deployment overhead. More broadly, \comic inherits limitations from its perception stack, including possible variation across languages, scripts, visual styles, and domains. Strong performance on English-centric benchmarks should therefore not be interpreted as universal robustness or fairness, and any deployment in sensitive settings should be accompanied by domain-specific validation and continued monitoring.


\section*{Acknowledgment}

The authors would like to thank...



%





\bibliographystyle{IEEEtran}
\bibliography{references}

\end{document}